\documentclass[epj]{svjour}

\usepackage{graphicx}
\usepackage{fancyhdr}
\def\makeheadbox{{
 }}
\def\mytitle{SO(10)-like Superstring Standard Model with R-parity
\\from the Heterotic String}
\def\myauthors{B. Kyae}
\def\mytype{Contributed Talk}
\def\mysession{Theoretical Models}

\begin{document}
\title{SO(10)-{\it like} Superstring Standard Model with R-parity
\\from the Heterotic String}

\author{Jihn E.  Kim,\inst{1}
 Ji-Hun Kim,\inst{1}
%
%
 \and
\underline{Bumseok Kyae}~ \inst{2}
 }  
 \institute{Department of Physics and Astronomy and Center for Theoretical
 Physics, Seoul National University, Seoul 151-747, Korea
 \and School of Physics, Korea Institute for Advanced Study, 207-43
Cheongryangri-dong, Dongdaemun-gu, Seoul 130-722, Korea}
%
\date{}
\abstract{We construct a supersymmetric standard model in the
context of the ${\bf Z}_{12-I}$ orbifold compactification of the
${\rm E_8\times E_8'}$ heterotic string theory. The gauge group is
${\rm SU(3)_c\times SU(2)_L\times U(1)_Y\times U(1)^4\times
[SO(10)\times U(1)^3]'}$ with ${\rm sin}^2\theta_W^0 = 3/8$. We
obtain three families of SO(10) spinor-like chiral matter states,
and Higgs doublets. All other extra states are exactly vector-like
under the standard model gauge symmetry. There are numerous
standard model singlets, many of which get VEVs such that only the
standard model gauge symmetry survives and desired Yukawa
couplings can be generated at lower energies. In particular, all
vector-like exotic states achieve superheavy masses and the
R-parity can be preserved.
} 

\maketitle
\section{Introduction}
\label{intro}

The standard model (SM) has been extremely successful in
explaining most particle physics phenomena of the scale larger
than $10^{-16}$ cm except gravity phenomena.
String theory is a promising candidate for a fundamental theory,
including quantum gravity as well as gauge interactions in the
framework.
%
%
However, the two approaches in theoretical physics, i.e. the
bottom-up and top-down approaches do not yet meet each other. It
would be an important task to connect the SM and string theory in
particle physics.
%

\noindent The minimal supersymmetric standard model (MSSM) is one
of the most promising candidates beyond the SM.  It would be an
effective theory appearing at an intermediate stage from the SM
toward string theory.
Thus, the features of the MSSM, which are listed below, are
expected to play an important role of the guide for a realistic
string model construction.

\begin{itemize}
\item[$\bullet$] Most of all, the gauge couplings in the MSSM, $g_3$,
$g_2$, $\sqrt{\frac{5}{3}}g_Y$ inferred from the RG evolutions are
unified at $10^{16}$ GeV energy scale. Thus, the ``${\rm
sin}^2\theta_W$'' (which is defined as $g_Y^2/[g_2^2+g_Y^2]$) is
$\frac{3}{8}$ at $10^{16}$ GeV.  It seems to imply the presence of
a unified interaction described by a grand unified theory (GUT) at
the $10^{16}$ scale. Particularly, the normalization factor of the
hypercharge, ``$\sqrt{\frac{5}{3}}$'' could not be easily
understood, were it not for GUTs such as SU(5) and SO(10)
theories. It remains undetermined within SM or MSSM, even if all
the gauge and gravity anomalies are considered.

\item[$\bullet$] One family of chiral matter in the MSSM, which are
16 chiral superfields $\{Q,d^c,u^c,L,e^c\nu^c\}$, are successfully
embedded in the multiplets of GUTs. In SU(5), for instance, they
are embedded in the the tensor (${\bf 10}$), (anti-)fundamental
(${\bf\overline{5}}$), and singlet ($\bf 1$) representations, and
in SO(10) embedded in a single spinor representation ($\bf 16$).
Namely, ${\bf 16}=(\pm\pm\pm\pm\pm)$ with even number of ``$-$''
[Throughout this article, ``$\pm$'' denotes $\pm\frac{1}{2}$.]
splits into ${\bf \overline{5}}_{-3}+{\bf 10}_1+{\bf 1}_5$ under
${\rm SU(5)\times U(1)_X}$, and
\begin{eqnarray}
{\bf \overline{5}}_{-3}&=&\bigg\{
\begin{array}{l}
(\underline{+--};--) \quad d^c
\\
(---;\underline{+-}) \quad L
\end{array} ,
\nonumber \\
{\bf 10}_1&=&\Bigg\{
\begin{array}{l}
(\underline{+--};++) \quad u^c
\\
(\underline{++-};\underline{+-}) \quad Q
 \\
(+++;--) \quad e^c
\end{array} ,
\label{16spinor} \\
{\bf 1}_{5}&=& ~~~ (+++;++) \quad \nu^c ,
\nonumber
\end{eqnarray}
where the underlined entries allow permutations.

\item[$\bullet$] More matter fields, if exist, composing SU(5)
multiplets leave intact the gauge coupling unification. Thus, the
value of ${\rm sin}^2\theta_W$ at the electroweak scale ($\approx
0.23$) could be a naturally predicted one in such GUTs.
\end{itemize}
The above features appearing in the MSSM seem to support the
presence of GUT.
The other aspects in the MSSM, however, make us doubtful of GUT.
\begin{itemize}
\item[$\bullet$] Unlike in chiral matter sector, the MSSM Higgs doublets
are not well-embedded in GUT multiplets unless unwanted triplet
partners ${\bf\overline{3}}$, ${\bf 3}$ are accompanied with them:
$\{H_d,{\bf\overline{3}}\}\subset {\bf\overline{5}}$, $\{H_u,{\bf
3}\}\subset {\bf 5}$ in SU(5), and ${\bf\overline{5}}$, ${\bf 5}$
are embedded in the vector representation ${\bf 10}$ in SO(10),
i.e. ${\bf 10}=(\underline{\pm
1,0,0,0,0})={\bf\overline{5}}_{2}+{\bf 5}_{-2}$ under ${\rm
SU(5)\times U(1)_X}$, and
\begin{eqnarray}
{\bf \overline{5}}_{2}&=&\bigg\{
\begin{array}{l}
(0,0,0;\underline{1,0}) ~~\quad H_d
\\
(\underline{1,0,0};0,0) ~~\quad {\bf\overline{3}}
\end{array} ,
\nonumber \\
{\bf 5}_{-2}&=&\bigg\{
\begin{array}{l}
(0,0,0;\underline{-1,0}) \quad H_u
\\
(\underline{-1,0,0};0,0) \quad {\bf 3}
\end{array}
\label{10vector} .
\end{eqnarray}
If the triplets ${\bf\overline{3}}$, ${\bf 3}$ remained light, the
gauge coupling unification in the MSSM would be destroyed.
However, making the triplets superheavy while keeping Higgs
doublets light is indeed non-trivial. The doublet-triplet
splitting is known to be a notorious problem in GUT.
\item[$\bullet$] In a simple SU(5) or SO(10), the masses of $d$-type
quarks and charged leptons should be the same, because $d$-type
quark Yukawa couplings, $Q_iH_dd^c_j$, and those of charged
leptons, $L_iH_de^c_j$ are unified in such GUTs: They are included
in ${\bf 10}_i{\bf \overline{5}}_j{\bf \overline{5}}_H$, in SU(5),
and in ${\bf 16}_i{\bf 16}_j{\bf 10}_H$ in SO(10). However, the
resulting mass relation $m_d=m_{e}^T$ is not realistic at all
except for the bottom and tau masses. It is difficult to avoid the
relation in GUT. In many GUT models, some vector-like pairs of
heavy fields and additional adjoint Higgs are introduced to
construct realistic Yukawa textures.
\end{itemize}
So far we have surveyed various features of the MSSM. Some aspects
in the MSSM might imply embedding of it in a GUT, but other
aspects seem to require that the MSSM should be rigorously just
SM-like.

Let us define the ``SO(10)-like MSSM,'' which reflects the
features discussed above. The SO(10)-like MSSM has the following
properties:
\begin{itemize}
\item[$\bullet$] The gauge group is just the SM gauge group,

${\rm SU(3)_3\times SU(2)_L\times U(1)_Y}$.
\item[$\bullet$] But the
${\rm sin}^2\theta_W$ is $\frac{3}{8}$.  It is a GUT property.
\item[$\bullet$] Three families of chiral matter fields are contained
in the three SO(10) spinor representations.
\item[$\bullet$] The two Higgs doublets are contained in the SO(10)
vector representation but the triplets are decoupled due to
asymmetry between doublets and triplets.
\item[$\bullet$] The $d$-type quark and charged lepton Yukawa
couplings are not summarized in a single Yukawa couplings,
avoiding $m_d=m_{e}^T$.
\item[$\bullet$] For stability of the proton and LSP, the R-parity (or
matter parity) is necessary.
\end{itemize}
We will suggest a model based on the heterotic string theory,
realizing the SO(10)-like MSSM.  The heterotic string theory
provides a good framework of unification, and its structure is
rich enough to accommodate the MSSM.  The heterotic string theory
is defined in 10D space-time, and its gauge group is ${\rm
E_8\times E_8'}$. In order to reduce the number of space
dimensions and also break gauge and supersymmetry (SUSY), we will
employ the orbifold compactification. The orbifold
compactification is relatively simple, and so easy to discuss
resulting low energy physics. Moreover, in orbifold
compactification, conformal field theory is still valid, and so
provides useful tools for analysis.

\section{The Model}

We employ the ${\bf Z}_{12-I}$ orbifold compactification, which is
specified with the twist vector, $\phi= \textstyle
(\frac{-5}{12}~\frac{4}{12}~\frac{1}{12})$.  It is associated with
the boundary conditions of the strings in the compact 6D space.
This twist vector leads to $N=1$ SUSY in the 4D space-time. For
the boundary conditions in the gauge coordinates, we take the
following form of a shift vector $V$ and a Wilson line
$a_3$~\cite{stringMSSM}:
\begin{eqnarray}
&&V=\textstyle (\frac14~\frac14~\frac14~\frac14~\frac14~
;\frac{5}{12}~\frac{5}{12}~\frac{1}{12})
(\frac14~\frac34~0~;0^5)' , \label{Z12Imodel} \nonumber \\
&&a_3=\textstyle
(\frac{2}{3}~\frac{2}{3}~\frac{2}{3}~\frac{-2}{3}~\frac{-2}{3}~;
\frac{2}{3}~0~\frac{2}{3}~)(0~\frac{2}{3}~\frac{2}{3}~;0^5)'\nonumber
.
\end{eqnarray}
They satisfy all the conditions required for modular
invariance~\cite{ChoiKimBk,flippedSU5}; $12(V^2-\phi^2)=12$,
$12a_3^2=48$, $12V\cdot a_3=12$. Since ``$\frac{1}{4}$''s are
aligned in the first five components in $V$, the visible gauge
symmetry is expected to be broken into SU(5). Due to the asymmetry
between the first three ($\frac{2}{3}$) and the next two entries
($\frac{-2}{3}$) in $a_3$, SU(5) would be further broken to ${\rm
SU(3)\times SU(2)}$. From the above forms of $V$ and $a_3$, ${\rm
SO(10)}'$ is expected in the hidden sector. We will clearly see
them later.

Low energy field spectrum in a model is determined by (1) the
massless conditions and (2) the GSO projection. The massless
conditions for the left and right movers on the orbifold ${\bf
Z}_{12-I}$ are
\begin{eqnarray}
\label{massless} &&{\rm left\ movers}:\
\frac{(P+kV_f)^2}{2}+\sum_iN^L_i\tilde{\phi}_i -\tilde c_k=0 ,
\nonumber \\
&&{\rm right\ movers}:\
\frac{(s+k\phi)^2}{2}+\sum_iN^R_i\tilde{\phi}_i-c_k=0, \nonumber
\end{eqnarray}
where $k=0,1,2,\cdots,11$, $V_f=(V+m_fa_3)$ with $m_f=0,+1,-1$,
and $i$ runs over $\{1,2,3,\bar{1},\bar{2},\bar{3}\}$.   Here
$\tilde{\phi}_j\equiv k\phi_j$ mod $Z$ such that
$0<\tilde{\phi}_j\leq 1$, and $\tilde{\phi}_{\bar{j}}\equiv
-k\phi_j$ mod Z such that $0<\tilde{\phi}_{\bar{j}}\leq 1$.
$N^L_i$ and $N^R_i$ indicate oscillating numbers for the left and
right movers. $P$ and $s$ [$\equiv (s_0,\tilde{s})$] are the ${\rm
E_8\times E_8'}$ and SO(8) weight vectors, respectively. The
values of $\tilde{c}_k$, $c_k$ are found in
Refs.~\cite{ChoiKimBk,flippedSU5}.

The multiplicity for a given massless state is calculated with the
GSO projector in the ${\bf Z}_{12-I}$ orbifold,
\begin{eqnarray} \label{phase}
{\mathcal P}_k(f) = \frac{1}{12\cdot 3}\sum_{l = 0 }^{11}
\tilde{\chi} ( \theta^k , \theta^l ) e^{2 \pi i l\Theta_k} ,
\nonumber
\end{eqnarray}
where $f$ $(=\{f_0, f_+, f_-\})$ denotes twist sectors associated
with $kV_f=kV$, $k(V+a_3)$, $k(V-a_3)$. The phase $\Theta_k$ is
given by
\begin{eqnarray}
&&\textstyle \Theta_k = \sum_i (N^L_i - N^R_i ) \hat{\phi}_i
+ (P+\frac{k}{2}V_f) V_f
 -(\tilde{s} + \frac{k}{2} \phi) \phi ,
 \nonumber
\end{eqnarray}
where $\hat{\phi}_j = \phi_{j}$ and
$\hat{\phi}_{\bar{j}}=-\phi_j$. Here, $\tilde{\chi}(\theta^k,
\theta^l)$ is the degeneracy factor summarized in
Ref.~\cite{flippedSU5}. Note that ${\cal P}_k(f_0)={\cal
P}_k(f_+)={\cal P}_k(f_-)$ for $k=0,3,6,9$.

In addition, the left moving states should satisfy
\begin{eqnarray}\label{condi3}
P\cdot a_3=0~~{\rm mod}~~{\rm Z}~~{\rm
in~the}~U,~T_3,~T_6,~T_9~{\rm sectors}. \nonumber
\end{eqnarray}

\subsection{Gauge symmetry and Weak mixing angle}

The untwisted sector ($k=0$) contains the gauge and chiral
multiplets as well as the gravity multiplet. The gauge group and
gauge quantum numbers are determined from the massless left mover
states. Since $\tilde{s}\cdot \phi=0$ for the gauge and gaugino
fields, only gauge multiplets satisfying $P\cdot V={\rm integer}$
and $P\cdot a_3={\rm integer}$ survives the projection with
$\Theta_0=0$ mod integer. The root vectors of ${\rm E_8\times
E_8'}$ satisfying $P\cdot V={\rm integer}$ and $P\cdot a_3={\rm
integer}$ in this model are only
\begin{eqnarray}
&&(\underline{1,-1,0};0,0,0^3)(0^8)',\quad
(0,0,0;\underline{1,-1};0^8)(0^8)',
\nonumber \\
&&\quad\quad\quad (0^8)(0^3;\underline{\pm 1,\pm1,0,0,0})' .
\label{roots}
\end{eqnarray}
They are the root vectors of SU(3), SU(2), and SO(10).
Since orbifold compactification preserves the rank of ${\rm
E_8\times E_8'}$, the gauge group of the model is
\begin{equation} \label{gaugegroup}
{\rm SU(3)_c\times SU(2)_L\times  U(1)_Y\times U(1)^4\times[SO(10)
\times U(1)^3]'} .
\end{equation}
%
%
In this model, ${\rm U(1)_Y}$ is defined using only the first five
components again:
\begin{eqnarray}
Y=\textstyle(\frac13~\frac13~\frac13~;~\frac{-1}{2}~
\frac{-1}{2}~;~0^3)(0^8)' , \nonumber
\end{eqnarray}
which is orthogonal to all non-Abelian root vectors in
Eq.~(\ref{roots}).  Thus, the SM quantum numbers of states are
read only from the first five components of $P+kV_f$:
\begin{eqnarray}
P+kV_f = ({\rm {\bf SM};x,x,x})({\rm x,x,x,x,x,x,x,x})' .
\nonumber
\end{eqnarray}
We will discuss later how the other visible gauge symmetries
except the SM gauge symmetry in Eq.~(\ref{gaugegroup}) can be
broken.

The current algebra in the heterotic string theory fixes the
normalization of $Y$ such that ${\rm U(1)_Y}$ is embedded in it.
Let us consider a properly normalized ${\mathbf Y}$:
\begin{eqnarray}
\textstyle {\mathbf Y} = u\times Y=
u\times\left[\sqrt{\frac{2}{3}}~\frac{\vec{q}_3}{\sqrt{2}}
-\frac{\vec{q}_2}{\sqrt{2}}\right] , \nonumber
\end{eqnarray}
where $u$ indicates a normalization factor of the hypercharge $Y$,
and $\vec{q}_3$
 and $\vec{q}_2$ are orthonormal bases:
$\vec{q}_3=\frac{1}{\sqrt{3}}(1,1,1;0,0;0^3)(0^8)'$ and
$\vec{q}_2=\frac{1}{\sqrt{2}}(0,0,0;1,1;0^3)(0^8)'$. For ${\mathbf
Y}$ to be embedded in the heterotic string theory, $u$ should be
fixed such that $u^2(\frac{2}{3}+1)=1$ or $u^2=\frac35$
~\cite{ChoiKimBk}. This hypercharge normalization leads to the
gauge coupling normalization $g_1^2=\frac53g_Y^2$, where $g_1$ is
unified at the string scale with the other non-Abelian gauge
couplings such as the ${\rm SU(2)_L}$ gauge coupling $g_2$. Thus,
in this model the weak mixing angle at the string scale is given
by
\begin{eqnarray}
 \sin^2\theta_W^0=\frac{1}{1+({g_2^2}/{g_Y^2})}
 =\textstyle\frac38 .
 \nonumber
\end{eqnarray}
It reflects a GUT property that this model has.

However, there is a considerable discrepancy between the string
and the GUT scales, $\Lambda_{\rm string}/\Lambda_{\rm GUT}\sim
10$. It could be avoided by simply assuming the heterotic
M-theory, in which $\Lambda_{\rm string}$ can be lowered to
$\Lambda_{\rm GUT}$. As another possibility, one could assume the
relatively large 6D space $\sim 1/\Lambda_{\rm GUT}$. Then the SM
gauge group can be embedded in a simple group SU(8) at the GUT
scale, which protects ${\rm sin}^2\theta_W=\frac{3}{8}$ up to the
string scale.

\subsection{Chiral matter}

The polarizations of matter states in the untwisted sector are in
the directions of the internal 6D space, and so $\tilde{s}\cdot
\phi$ for the states with the left handed chirality ($\chi ={\rm
L}$) is $\frac{-5}{12}$, $\frac{4}{12}$, or $\frac{1}{12}$ mod Z.
For $\Theta_0=0$ mod Z, $P\cdot V$ should be one of
$\{\frac{-5}{12},
 \frac{4}{12}, \frac{1}{12}\}$ mod Z. The ${\rm E_8\times E_8'}$
root vectors satisfying it and $P\cdot a_3={\rm integer}$ are
listed in {\bf Table~1}.
%
%
\begin{table}
\caption{Matter states from the $U$ sector} \label{tab:1}
\begin{tabular}{lllll}
\hline\noalign{\smallskip}
$P\cdot V$ & States ($P$) & $\chi$ & SM & $\Gamma$   \\
\noalign{\smallskip}\hline\noalign{\smallskip} $\frac{-5}{12}$ &
$(\underline{++-};\underline{+-};+++)(0^8)'$ & L
& $Q_3$ & +1 \\
($U_1$) & $(---;\underline{+-};+--)(0^8)'$ & L & $L_3$ & $-3$
\\
\noalign{\smallskip} \hline \noalign{\smallskip}
$$ & $(\underline{+--};--;+++)(0^8)'$ & L & $d^c_3$ & $+1$
\\
$\frac{1}{12}$ & $(+++;++;+++)(0^8)'$ & L &$\nu^c_3$ & $+1$
\\
($U_3$) & $(\underline{+--};++;+--)(0^8)'$ & L & $u^c_3$ & $-3$
\\
$$  & $(+++;--;-+-)(0^8)'$ & L & $e^c_3$ & $+5$ \\
\noalign{\smallskip}\hline \noalign{\smallskip}
$$ & $(0,0,0;\underline{1,0};0,0,1)(0^8)'$ &
L & $H_d$ & $-2$
\\
$\frac{4}{12}$
($U_2$) & $(0,0,0;\underline{-1,0};-1,0,0)(0^8)'$ & L & $H_u$ &
$+2$
\\
$$ & $(0,0,0;0,0;1,0,-1)(0^8)'$ & L & ${\bf
1}_0$ & $0$
\\
\noalign{\smallskip}\hline
\end{tabular}
\vspace*{0.5cm}
\end{table}
%
%
Note that the first five entries in $P$ of the states from the
$U_1$ and $U_3$ sectors take {\it exactly the form of SO(10)
spinor} ${\bf 16}$ shown in Eq.~(\ref{16spinor}). Thus, we get one
family of the MSSM chiral matter from the untwisted sector. From
the $U_2$ sector, we have the MSSM Higgs doublets and a SM
singlet. The Higgs doublets are {\it pieces of an SO(10) vector}
shown in Eq.~(\ref{10vector}). However, the unwanted triplets are
absent in the untwisted sector, because they can not satisfy the
projection conditions.

We need two more families of the MSSM chiral matter. We find them
from the $T_4^0$ twist sector. The $T_4^0$ sector is the sector
constructed with the string states satisfying the boundary
conditions, $4\times\phi$ and $4\times V^I$. The gauge quantum
numbers ($P+4V$) of some massless states and the numbers of the
corresponding states determined with ${\cal P}_4$ are listed in
{\bf Table~2}.
%
%
\begin{table}
\caption{Some matter states from the $T_4^0$ sector} \label{tab:1}
\begin{tabular}{lllll}
\hline\noalign{\smallskip}
States ($P+4V$) & $\chi$ & ${\cal P}_4$ & SM & $\Gamma$   \\
\noalign{\smallskip}\hline\noalign{\smallskip}
$\left(\underline{+--};--;\frac{1}{6},\frac{1}{6},\frac{-1}{6}\right)
(0^8)'$ & L & $2$ &$2\cdot d^c$ & $+1$
\\
$\left(---;\underline{+-};\frac{1}{6},\frac{1}{6},\frac{-1}{6}\right)
(0^8)'$ & L & $2$ & $2\cdot L$ & $-3$
\\
$\left(\underline{+--}++;\frac{1}{6},\frac{1}{6},\frac{-1}{6}\right)
(0^8)'$ & L & $2$ & $2\cdot u^c$ & $-3$
\\
$\left(\underline{++-};\underline{+-};
\frac{1}{6},\frac{1}{6},\frac{-1}{6}\right) (0^8)'$ & L & $2$ &
$2\cdot Q$ & $+1$
\\
$\left(+++;--;\frac{1}{6},\frac{1}{6},\frac{-1}{6}\right) (0^8)'$
& L & $2$ & $2\cdot e^c$ & $+5$
\\
$\left(+++;++;\frac{1}{6},\frac{1}{6},\frac{-1}{6}\right) (0^8)'$
& L & $2$ & $2\cdot \nu^c$  & $+1$
\\
\noalign{\smallskip}\hline
\end{tabular}
\vspace*{0.5cm}
\end{table}
%
%
We see again the {\it SO(10) spinor's structure} from the first
five entries in $P+4V$. They are identified as two copies of ${\bf
16}$s. Hence, including one family of ${\bf 16}$ from the
untwisted sector, we have in total three families of ${\bf 16}$s.
%
%
Of course, there are more matter states in $T_4^0$ and a lot of
massless states including SM singlets also arise in other sectors.
They all should be counted for modular invariance of the model.
However, all the other matter except the MSSM chiral fields found
in the above turn out to be {\it exactly vector-like under the SM
gauge symmetry}~\cite{stringMSSM}.

In Ref.~\cite{stringMSSM}, all gauge and gravity anomalies in this
model have been checked out. Only one ${\rm U(1)_A}$ symmetry
defined with $Q_A=(-6^5;1,1,1)(1,-1;0^6)'$ turns out to be
anomalous. The anomaly could be cancelled via the Green-Schwartz
mechanism. It is a general feature in the heterotic string theory.

\subsection{Yukawa couplings and Desirable vacuum}

In order to discuss Yukawa couplings and desired vacuum, we need
to know the allowed superpotential in the string model.
The superpotential are obtained from vertex operators obeying the
orbifold conditions~\cite{ChoiKimBk}. They are summarized as the
following selection rules:
\begin{itemize}
\item[(a)] Gauge invariance.
\item[(b)] the $H$-momentum, which is defined by
$R_j=(\tilde{s}+k\phi+\tilde{r}_-)_j-N^L_j+N^L_{\bar{j}}$, should
be preserved:
\begin{eqnarray}
&&\textstyle \sum_z R_1 (z) = 1 {\rm~ mod~} 12 , \quad \sum_z R_2
(z) = 1 {\rm~ mod~} 3,
\nonumber \\
&&\textstyle \quad\quad\quad\quad\quad \sum_z R_3 (z) = 1 {\rm~
mod~} 12,\label{Hconsv} \nonumber
\end{eqnarray}
where $z$ denotes the index of states participating in a vertex
operator. The $H$-momentum conservation is a remnant of the
Lorentz symmetry of the internal 6D space, and interpreted as a
discrete R-symmetry in the 4D space-time.
\item[(c)] Since a discrete symmetry associated with
${\rm {\bf Z}}_{12-I}$ has been introduced, the space group
selection rules should be satisfied:
\begin{eqnarray}
\textstyle \sum_z k(z) = 0 {\rm~ mod~} 12,\quad \label{modinvk}
\sum_z \left[ km_f \right] (z) = 0 {\rm~ mod~} 3.\label{modinva}
\nonumber
\end{eqnarray}
\end{itemize}
In this model, the allowed renormalizable Yukawa couplings from
the untwisted sector are only
\begin{eqnarray}
U~{\rm sector}~:~~Q_3H_uu^c_3~,\quad L_3H_de^c_3 ~, \quad
L_3H_u\nu^c_3  ~. \nonumber
\end{eqnarray}
But the R-parity violating couplings are forbidden by the
selection rules. The above allowed Yukawa couplings would be
phenomenologically quite acceptable with large ${\rm tan}\beta$ ($
\sim 50$) except for the missing bottom quark Yukawa coupling. The
Yukawa couplings in the untwisted sector are from the 10D gauge
couplings.
%
%
Once the SM singlets develop VEVs of order $\Lambda_{\rm string}$,
however, the above discussion might not be so valid.

The superpotential includes the terms constructed purely with the
SM singlets. In ${\rm {\bf Z}}_{12-I}$ orbifold compactification,
if a superpotential term $\omega$ satisfies all the selection
rules, higher order terms $\omega^{12n+1}$ ($n=1,2,3,\cdots$) also
do. By including $\omega^{13}$, $\omega^{25}$, $\cdots$ in the
superpotential,  $W\supset \omega\left[1+\omega^{12}/\Lambda_{\rm
string}^{36}+\cdots\right]$, one can always find a vacuum with
$\langle\omega\rangle\sim\Lambda_{\rm string}^3$. Thus, the
singlets of interest can develop VEVs of the string scale,
preserving all the F-flat conditions $F_i^*=D_iW=\frac{\partial
W}{\partial\phi^i}+\frac{\partial
K}{\partial\phi^i}\frac{W}{M_P^2}=0$, where $\phi^i$ denotes the
SM singlets.

In fact, the ``no-scale'' structure in supergravity is easily
broken (e.g. due to the modulus dependence of the superpotential).
Then the F-flat solution does not mean $\langle W\rangle =0$ in
general. Then {\it all} the D-flat conditions including that of
${\rm U(1)_A}$ can be also fulfilled:
$D^a=gG_{i}T^a_{ij}\phi^j=g\frac{M_P^2}{W}D_iWT^a_{ij}\phi^j=0$~\cite{KKLT}.

Let us assume that SM singlets develops VEVs, $\langle {\bf
1}_0\rangle{\rm s} \sim \Lambda_{\rm string}$.  Since they could
carry non-zero charges of some U(1)s unobserved at low energies,
the U(1) factors in Eq.~(\ref{gaugegroup}) can be broken while the
SM gauge symmetry still preserved. From couplings of $\langle{\bf
1}_0{\bf 1}_0'\cdots\rangle\Phi\bar{\Phi}$, unwanted vector-like
pairs of exotics (or extra fields unobserved at low energies)
achieve superheavy masses. Even if we didn't achieve all the
needed SM Yukawa couplings at the renormalizable level, we can
obtain them from nonrenormalizable couplings with $\langle {\bf
1}_0\rangle {\rm s}\sim \Lambda_{\rm string}$. However, unless the
R-parity survives down to low energies, the R-parity violating
terms would be also induced with couplings of order unity, because
basically they also respect the SM gauge symmetry. Thus, on a
phenomenologically desirable SUSY vacuum with $\langle {\bf
1}_0\rangle{\rm s}\sim \Lambda_{\rm string}$,
\begin{itemize}
\item[$\bullet$] only SM gauge symmetry should survive at low energies,
\item[$\bullet$] all unwanted exotics should achieve heavy
masses and be decoupled from low energy physics, and
\item[$\bullet$] all the desired SM Yukawa couplings should be induced
while the R-parity violating terms absent (or
suppressed)~\cite{stringMSSM,flippedR}.
\end{itemize}
In fact, many trials to make all exotics heavy by large $\langle
{\bf 1 }_0\rangle$ end up with R-parity breaking. In this model,
however, it is possible to separate the SM singlets into two (or
more) classes I and II such that the requirements listed above are
fulfilled~\cite{stringMSSM}: The SM singlets belonging to the
class I develop VEVs $\sim\Lambda_{\rm string}$, by which all the
exotics achieve superheavy masses and all the needed SM Yukawa
couplings are induced {\it with leaving intact the R-parity}. On
the other hand, the VEVs of the singlets in the class II are
assumed to be zero (or small enough), which make it possible to
define the exact (or effective) R-parity.

\section{Conclusion}

Thus, based on the heterotic string theory compactified on the
${\bf Z}_{12-I}$ orbifold, we have successfully realized the
``SO(10)-like MSSM'' defined in Introduction.

\end{document}